\documentclass[useAMS,usenatbib,letterpaper]{mn2e}
\usepackage{url}
\usepackage{times}
\title[Galaxy Zoo]{Galaxy Zoo 1 : Data Release of Morphological Classifications for nearly 900,000 galaxies\thanks{This publication has been made possible by the participation of more than 100,000 volunteers in the Galaxy Zoo project. Their contributions are individually acknowledged at http://www.galaxyzoo.org/Volunteers.aspx}}
\author[Lintott et al.]{
\parbox[t]{16cm}{Chris Lintott$^{1,2}$\thanks{Email: cjl@astro.ox.ac.uk}, Kevin Schawinski$^{3,4}$\thanks{Email: kevin.schawinski@yale.edu},  Steven Bamford$^{5}$, An\v{z}e Slosar$^{6}$,  Kate Land$^{1}$, Daniel Thomas$^{7}$,  Edd Edmondson$^{7}$,Karen Masters$^{7}$, Robert C. Nichol$^{7}$,M. Jordan Raddick$^{8}$, Alex Szalay$^{8}$,Dan Andreescu$^{9}$, Phil Murray$^{10}$, Jan Vandenberg$^{8}$\\}\\ 
$^{1}$Oxford Astrophysics, Denys Wilkinson Building, Keble Road, Oxford, OX1 3RH, UK\\
$^{2}$Adler Planetarium, 1300 S. Lake Shore Drive, Chicago, IL 60605, U.S.A.\\
$^{3}$Einstein Fellow, Department of Physics, Yale University, New Haven, CT 06511, U.S.A.\\
$^{4}$Yale Center for Astronomy and Astrophysics, Yale University, P.O. Box 208121, New Haven, CT 06520, U.S.A.\\
$^{5}$Centre for Astronomy \& Particle Theory, University of Nottingham, University Park, Nottingham, NG7 2RD, UK\\
$^{6}$Brookhaven National Laboratory, Upton NY 11973, USA\\
$^{7}$Institute of Cosmology and Gravitation, University of Portsmouth, Dennis Sciama Building, Burnaby Road, Portsmouth, PO1 3FX, UK\\
$^{8}$Department of Physics and Astronomy, Johns Hopkins University, 3400 N. Charles St., Baltimore, MD 21218, USA\\
$^{9}$LinkLab, 4506 Graystone Ave., Bronx, NY 10471, USA\\
$^{10}$Fingerprint Digital Media, 9 Victoria Close, Newtownards, Co. Down, Northern Ireland, BT23 7GY, UK\\
}

\usepackage[dvips]{graphicx}
\usepackage{setspace}

\usepackage{deluxetable}

\begin{document}

\date{July 2010}
\pagerange{\pageref{firstpage}--\pageref{lastpage}} \pubyear{2010}

\maketitle

\label{firstpage}

\begin{abstract}

Morphology is a powerful indicator of a galaxy's dynamical and merger history. It is strongly correlated with many physical parameters, including mass, star formation history and the distribution of mass. The Galaxy Zoo project collected simple morphological classifications of nearly 900,000 galaxies drawn from the Sloan Digital Sky Survey, contributed by hundreds of thousands of volunteers. This large number of classifications allows us to exclude classifier error, and measure the influence of subtle biases inherent in morphological classification. This paper presents the data collected by the project, alongside measures of classification accuracy and bias. The data are now publicly available and full catalogues can be downloaded in electronic format from \url{http://data.galaxyzoo.org}.

\end{abstract}


\section{Introduction}

The aim of the Galaxy Zoo project was to provide visual morphologies for nearly one million galaxies from the Sloan Digital Sky Survey \citep{York}, including the whole Main Galaxy Sample (MGS) \citep{Strauss}. Separating galaxies into categories based on their morphology (the visual appearance or shape) has been standard practice since it was first systematically applied by \citet{Hubble}. These morphological categories are broadly correlated with other, physical parameters, such as the star formation rate and history, the gas fraction and dynamical state of the system \citep{Roberts}; understanding these correlations and studying the cases where they do not apply is critical to our understanding of the formation and evolution of the galaxy population. It is tempting to identify the morphological distinctions with the clear colour bimodality in the population of galaxies visible in data from modern surveys (e.g \citealt{Strateva}), but extremely large sets of classified galaxies are necessary before this hypothesis can be tested. For most of the twentieth century, morphological catalogues were compiled by individuals or small teams of astronomers (e.g. \citealp{Sandage,deVaucouleurs}), but modern surveys containing many hundreds of thousands of galaxies make this approach impractical. 

Attempts to solve this problem have taken one of three approaches. The first is to use physical parameters, such as colour, concentration index, spectral features, surface brightness profile, structural features, spectral energy distribution \citep{Kinney} or some combination of these as a proxy for morphology (e.g. \citealp{Abraham, Conselice}). As no proxy is an exact substitute for true visual morphology, the introduction of each of these variables results in an unknown and potentially unquantifiable bias in the resulting sample of galaxies. Although morphological labels are often used for the resulting catalogues, usually after comparison with a small number of expert classifications, each of these techniques produces catalogues that cannot entirely reproduce true morphological selection. For example, \citet{Schawinski} showed that the proportion of elliptical galaxies with recent star formation or nuclear activity was significantly higher in a sample classified by eye than in samples assembled using proxies for morphology. This reflects that fact that proxy quantities such as colour do not directly probe the dynamical state of the system which controls the morphology.  

The second strategy was applied by \citet{Lahav} and then further developed by \citet{Ball} amongst others. The aim was to develop automatic classification routines, typically neural networks, to the point that they can replace the need for human classifications. While successful in classifying the majority of galaxies, a major problem is that due to both the use of proxies for morphology as input and the inherent complexity of the network, it is not easy to predict or understand the bias in the resulting classifications. As a result these automatic classifiers have not been widely adopted. 

The third approach is to attempt to expand the reach of visual classifications. Previous professional attempts \citep{Fukugita, Nair} have been necessarily limited by the extraordinary effort required to classify even relatively small subsets of the SDSS; the largest, MOSES \citep{Schawinski}, included basic classifications of only 50,000 galaxies at redshifts between 0.05 and 0.1, and with $r<16.8$~($\sim 5.5$ per cent of the SDSS). The results presented in this paper, which expands on our first description of the Galaxy Zoo project and its results \citep{Lintott08}, provide estimates of the visual morphology of the entire SDSS main galaxy sample. Having produced a catalogue of visual morphologies, we can use the other measured parameters, including colour, to investigate the galaxy population. 

Galaxy Zoo is possible because of the involvement of hundreds of thousands of volunteer `citizen scientists'. The involvement of non-professionals in astronomical science has a long and distinguished history. From the early contributions of observers to modern discoveries of supernov\ae~ and follow-up of candidate extra-solar planets (e.g. \citealt{Barbieri}), astronomers have often depended on volunteers. The Galaxy Zoo project expands the role of non-professionals in astrophysics from data collection to include data analysis, a technique was first successfully employed in astronomy or astrophysics by the Stardust@Home project \citep{Westphal, Mendez}. Its usefulness is demonstrated both by the catalogue presented here, but also by the serendipitous discovery of unusual objects and classes of object discussed elsewhere (e.g. \citealp{voorwerp,peas}). 

\section{Sample selection and web site operation}
\label{sec:sample}

The images of galaxies presented for classification by Galaxy Zoo were drawn from the Sloan Digital Sky Survey, a survey of a large part of the northern sky providing photometry in five filters \citep{Fukugita96,Smith} and covering $\sim26$ per cent of the sky. Data Release 6 (DR6, \citealt{DR6}) was used for the initial selection of candidates. We include the Main Galaxy Sample (MGS) of \citet{Strauss} which includes all extended objects with Petrosian magnitude $r<17.77$, a total of 738,175 galaxies including those for which spectra were not available at the time of the DR6 release. In order to be as inclusive as possible, 155,037 objects which had been included in the SDSS spectroscopic survey (for various reasons, including meeting criteria intended to select for luminous red galaxies, quasars and other unusual objects) and subsequently classified as a galaxy due to their spectral properties were added, making a total of 893,212 objects. 

\begin{figure}
\includegraphics[width=0.35\textwidth,angle=270]{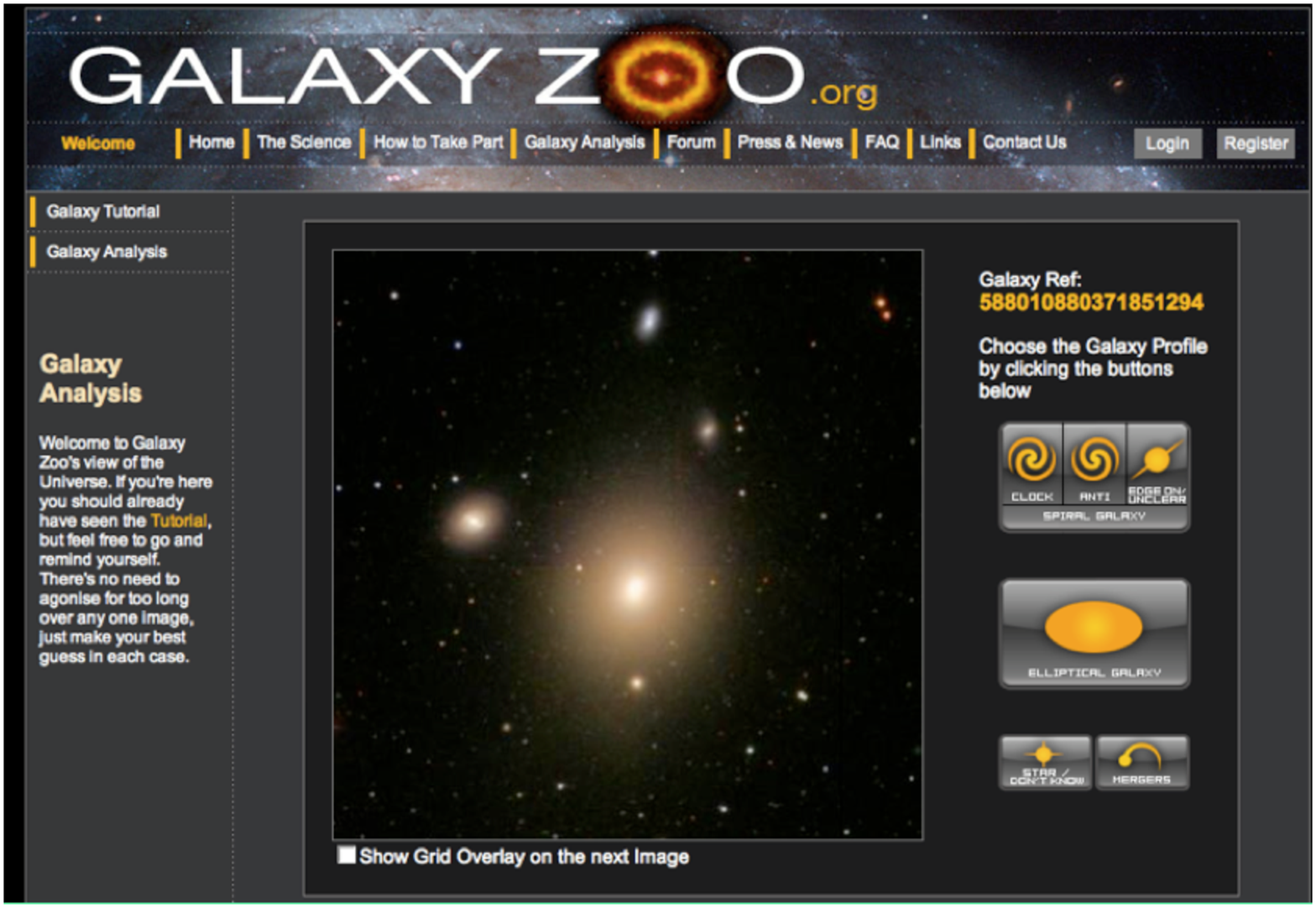}
\caption{Main analysis page from the Galaxy Zoo website.}\label{fig:screenshot}
\end{figure}

Composite images of these objects in the $g$,$r$ and $i$ bands were provided by the ImgCutout web service \citep{Nieto} on the SDSS SkyServer website \citep{Szalay02} and then shown to visitors to the Galaxy Zoo website\footnote{The original Galaxy Zoo website is maintained and archived at \url{http://zoo1.galaxyzoo.org}}, who were then asked to classify the galaxy into one of the six categories shown in Table \ref{tab:buttons}. Distinguishing clockwise and anticlockwise spirals was not only useful in its own right, but also allowed \citet{Masters2} to ensure that their sample of red spirals genuinely included only spirals and not edge-on disks or possible S0 galaxies; spiral arms must have been seen by a majority of classifiers to record significant evidence for either a clockwise or an anticlockwise spin. The size of the image of each galaxy was chosen so that the scale was 0.024$\rmn{R_p}$ arcsec per pixel, where $\rmn{R_p}$ is the Petrosian radius for the system. The images were $423$ pixels ($\approx 10 R_\rmn{p}$ for a typical system) on each side. The interface is shown in Figure \ref{fig:screenshot}. The web site was launched on 11th July 2007, and full details of its operation are given in \citet{Lintott08}. 

\begin{table}
\begin{tabular*}{0.45\textwidth}{@{\extracolsep{\fill}}ccl}
\hline
Class & Button & Description \\
\hline
1 & \raisebox{-0.5ex}{\includegraphics[width=0.5cm]{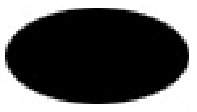}}
& Elliptical galaxy \\
2 & \raisebox{-2.0ex}{\includegraphics[width=0.5cm]{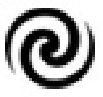}}
  & Clockwise/Z-wise spiral galaxy \\
3 & \raisebox{-2.0ex}{\includegraphics[width=0.5cm]{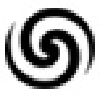}}
  & Anti-clockwise/S-wise spiral galaxy \\
4 & \raisebox{-2.0ex}{\includegraphics[width=0.5cm]{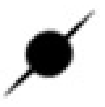}}
  & Spiral galaxy other (eg. edge on)\\
5 & \raisebox{-2ex}{\includegraphics[width=0.5cm]{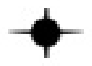}}
& Star or Don't Know (eg. artefact) \\
6 & \raisebox{-2ex}{\includegraphics[width=0.5cm]{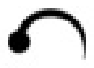}}
& Merger \\
\hline
\end{tabular*}
\caption{Galaxy Zoo classification categories showing schematic symbols as used on the site.}\label{tab:buttons}
\end{table}

\subsection{Data reduction}
\label{sec:prop}
The data reduction required to turn clicks provided via a website into a scientific catalogue is substantial. As well as comparing Galaxy Zoo with existing professional catalogues, \citet{Lintott08} explored the possibility of weighting users according to how often they agreed with the majority, but found little change to the resulting classifications. Requirements for 80 per cent and 95 per cent agreement amongst users were then used to define `clean' and `superclean' samples respectively. This approach, while suitable for some purposes, has proved to be inadequate for others. \citet{Darga}, in their study of merging galaxies, found that all galaxies with a fraction of 40 per cent or more of their vote in the `merger' category were, in fact, true mergers. This result suggests that the application of a single critical threshold to all classifications is over-simplistic. The catalogue presented in this paper therefore includes the fraction of clicks in each category for all galaxies, rather than just those in the `clean' or `superclean' samples. Users of the dataset presented here are, however, recommended to use cuts of 0.8 or 0.95 in the first instance to ensure where possible that results are comparable with earlier results. 

In using the Galaxy Zoo morphologies, it is important to consider the population of galaxies which are unclassified according to the criterion used to assign individual galaxies to a population. It is obviously possible to derive a classification for every galaxy by simply assigning it to the category with the greatest fraction of the vote (which we refer to as the \texttt{greater} criterion); a galaxy with 51 per cent of the vote in the elliptical category would, in this system, be considered an elliptical. For more stringent thresholds (e.g. \texttt{clean}, where a galaxy would require at least 80 per cent of the vote to be assigned a classification) then unclassified galaxies may form a majority of the sample. In order to evaluate the effect of this feature of the data, we determine the fraction of unclassified galaxies as a function of magnitude and size. 

In an effort to quantify the effect of other potential biases in the classification process, mirrored and greyscale images were introduced to the site from 28 November 2007. The greyscale images were not single filter images, but in order to minimize the effect of apparent changes in morphology caused by viewing the galaxy in different wavelengths were instead produced from the $gri$ colour images provided by the SDSS pipeline. A subsample of the main catalogue sample was used, comprising the \texttt{superclean} sample as of 4th September 2007 (i.e. all galaxies with an agreement of more than 95 percent on that date) and a random sampling of 5 percent of the rest of the sample, comprising 91,303 images in total. A list of galaxies included in this bias study sample is given in section \ref{sec:classifications}.

The results of this bias study were discussed in \citet{Lintott08} and \citet{Land}. Significant but small biases in spin direction and colour were found, with anticlockwise spiral classifications favoured over clockwise, and the greyscale images were more likely to be classified as elliptical than their colour counterparts. 

Although these biases were small, the effect of studying them was relatively large. Any study of the behaviour of human classifiers is likely to encounter a phenomenon known as the Hawthorne effect \citep{Mayo}, the risk of changing the behaviour of those taking part in a study simply by carrying out the study itself. A change in classifier behaviour was indeed observed, with users being slightly more careful in their classifications during the bias study and thus classifying fewer galaxies as spiral. The effect is small ($\sim 3$ per cent fewer votes were received in the spiral categories) but significant. Rather than just combining classifications for each of the galaxies included in the bias study, therefore, we present the data from before and then during the study separately. 

\section{Properties of the data}
\subsection{Quantifying bias}
\label{sec:bias}

\citet{Bamford} carry out a more sophisticated analysis of the Galaxy Zoo data, initially motivated by the desire to determine the relationship between morphology and the local density of galaxies. The technique used recognises that although small, faint or distant galaxies will likely appear as ellipticals and therefore will be classified as such with a high degree of agreement, many of these systems are likely to be spirals whose arms are not distinctly visible in the SDSS images. By assuming that the morphological fraction within bins of fixed galaxy size and luminosity does not evolve over the depth of the SDSS, it is possible to statistically estimate the bias affecting the morphological classifications for galaxies of known luminosity, size and distance. It is important to note that this bias does not arise from the involvement of volunteers in the classification process, but from the inherent limitations of the survey data.

The method only deals with removing the bias relative to the least biased data (i.e. that from nearby systems) and so there may be a remaining, unquantified bias, for example due to bias in human pattern recognition abilities. The bias correction will also reduce the impact of any true redshift evolution from the sample (although only that evolution which affects the morphological mix of the population at a given absolute magnitude and physical size). 

The effect of this bias on a final catalogue depends on how the raw Galaxy Zoo classifications are treated.  As an example, if a simple majority of the vote is used to classify galaxies then $\sim 13.5$ per cent of galaxies are in absolute magnitude--size bins where approximately no bias correction is necessary, and $\sim 20$ per cent are in bins which have approximately no misclassified galaxies.  Conversely, $\sim 6$ percent of galaxies are in bins where the bias correction changes the classifications for more than half of the objects. In contrast, if the \texttt{clean} criterion is used then a much larger fraction, over 70 per cent, of galaxies are in absolute magnitude size bins for which no objects are misclassified due to this bias. The price of applying a more stringent criterion for classification is thus a large fraction of objects which do not meet the \texttt{clean} criteria and thus have `uncertain' classifications.

As determining the bias correction requires a redshift, debiased results are only available for objects which were spectroscopically observed by SDSS, a subset of the whole Galaxy Zoo sample. The determination also requires a well-defined, homogeneous sample and is therefore limited to MGS objects with reliable $r$-band photometry, redshifts in the range $0.001$--$0.25$ and absolute magnitudes and sizes that are not extreme outliers from the normal galaxy distribution. \citet{Bamford} used DR6 of the SDSS which only provided spectroscopic coverage for 82 percent of the survey area.  With the availability of SDSS Data Release 7 \citep{Abazajian}, the spectroscopic coverage has risen.  As a result, the number of objects with redshifts has increased from 677,515 (76 percent) to 781,842 (88 percent).  Considering just the Main Galaxy Sample, the total number of objects in Galaxy Zoo is 738,173, of which 679,721 (92 percent) have redshifts in DR7, up from 575,398 (78 percent) in DR6.  In calculating the classification bias corrections we have thus supplemented our previous DR6 catalogue with additional redshifts from DR7, significantly increasing the fraction of the sample for which we can provide these corrections.

As described in appendix A of Bamford et al. 2009, we first divide the
sample into bins of similar luminosity, physical size and redshift. We then
find for each point in luminosity--size space the lowest redshift bin
containing at least 30 galaxies, and assume that this bin represents the
`true' early-type to spiral ratio.  In an attempt to keep this
baseline estimate unbiased, we only consider bins well
away from the magnitude, size and surface brightness limits of the sample.

Having obtained an approximation to the low-redshift early-type to spiral
ratio as a function of both luminosity and size, we fit an appropriate smooth function to the result. Equation A1 in Bamford et al. gives the fit

\begin{equation}
  \left< \frac{n_{\rmn{el}}}{n_{\rmn{sp}}} \right>_{\rmn{base}} = \frac{p_1}{1 + \exp\left(
      \frac{s_1(R_{50}) - M_r}{s_2(R_{50})}\right)} + p_2,
\end{equation}
where,
\begin{eqnarray}
s_1(R_{50}) &=& q_1^{-(q_2 + q_3 {R_{50}}^{q_4})} + q_5,\nonumber \\
\rmn{and} &&\\
s_2(R_{50}) &=& r_1 + r_2 (s_1(R_{50}) - q_5). \nonumber
\end{eqnarray}

By considering the difference between this baseline
early-type to spiral ratio and that measured for a given bin
of absolute magnitude, physical size and redshift, we can estimate the
correction, $C$, required for any particular galaxy as 
\begin{equation}
C(M_r, R_{50}, z) = \log \left(\frac{\left< n_{\rmn{el}}/n_{\rmn{sp}}
    \right>_{\rmn{raw}}}{\left< n_{\rmn{el}}/n_{\rmn{sp}} \right>_{\rmn{base}}}\right),
\end{equation}
where angular brackets indicate averages over bins of ($M_r$, $R_{50}$, $z$).

Individual vote shares (type likelihoods) for each galaxy are adjusted as
\begin{eqnarray}
p_{\rmn{el,adj}} &=& \frac{1}{1\big/\big(\frac{p_{\rmn{el}}}{p_{\rmn{sp}}}\big)_{\rmn{adj}} + \frac{p_x}{p_{\rmn{el}}} + 1},\\
p_{\rmn{sp,adj}} &=& \frac{1}{\big(\frac{p_{\rmn{el}}}{p_{\rmn{sp}}}\big)_{\rmn{adj}} + \frac{p_x}{p_{\rmn{sp}}} + 1}, \nonumber
\end{eqnarray}
where
\begin{equation}
{\scriptstyle\left(\frac{p_{\rmn{el}}}{p_{\rmn{sp}}}\right)_{\rmn{\!adj}}} = {\scriptstyle\left(\frac{p_{\rmn{el}}}{p_{\rmn{sp}}}\right)_{\rmn{\!raw}}} \Big/ 10^{C(M_r, R_{50}, z)}, \nonumber
\end{equation}
and
$p_x = 1 - p_{\rmn{el}} - p_{\rmn{sp}}$.

\begin{figure*}
\centering
\includegraphics[height=\textwidth,angle=270]{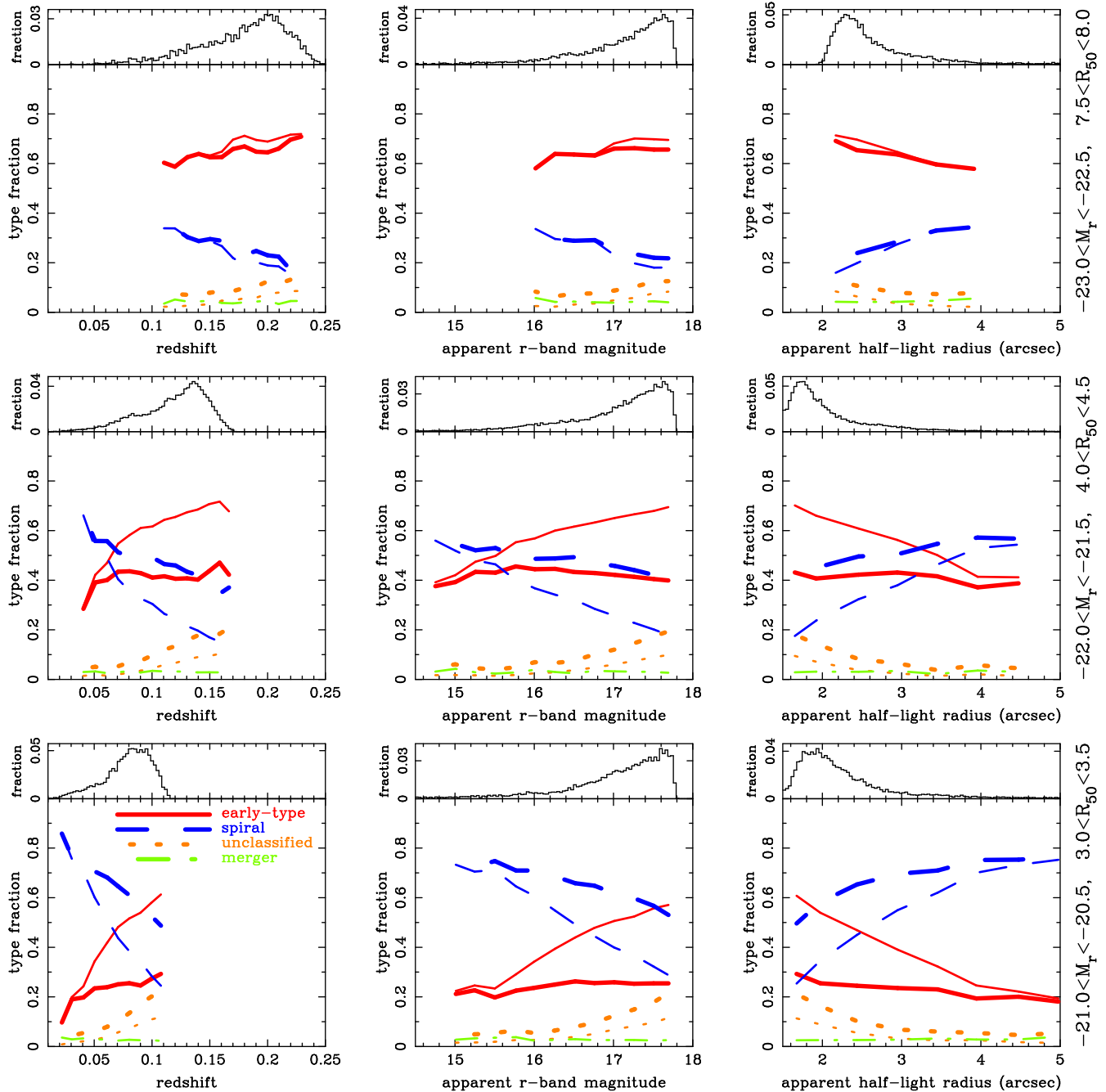}
\caption{\label{fig:addbias}
The effect of the bias, and of the result of adopting the measured correction as a function of redshift, apparent magnitude and apparent size. The thin and thick lines correspond to de-biased and raw classifications respectively.}
\end{figure*}

Figure \ref{fig:addbias} illustrates the effect of the bias, and the result of
adopting the measured correction, as a function of redshift, apparent magnitude
and apparent size, for three bins of absolute magnitude and
physical size.  The overall result of applying the correction is to lower the
probability that a given galaxy will be classified as early-type and
increase the chance that it will classified as spiral. The effect
is largest around the median SDSS redshift, and for faint, small
galaxies; bright, large, low redshift galaxies
need only a small correction, whereas at the highest redshifts only
the most luminous galaxies pass the sample selection criteria, most of
which are indeed early-types.

\subsection{Measures of confidence}

\label{sec:confidence}
To assist users of the Galaxy Zoo dataset in evaluating the morphological classifications they obtain, both individually and for larger samples, we have calculated a number of relevant statistics. These were derived from the bias correction procedure described above and thus reflect only the sensitivity of the classification to the bias described in the previous section. They do not take into account other systematic biases that may exist in the data set (see \citealp{Lintott08} for a comparison of the Galaxy Zoo classifications with other catalogues of visual morphology). 

A first indicator of the quality of an individual morphology is the difference between its raw and debiased likelihoods, $\Delta p$. The bias corrections are inherently uncertain, especially when applied to individual galaxies rather than to large samples, but the size of the bias correction is an indication of the uncertainty in the galaxy's type.

To remove the individual uncertainties on our confidence measures, for each galaxy we calculate values computed from a `bin' of galaxies with similar redshift, absolute magnitude and physical size, corresponding to the same binning used in quantifying the bias correction.  We therefore provide the mean and standard deviation of $\Delta p$ in each galaxy's bin, $\left< \Delta p \right>$ and $\sigma_{\Delta p}$ respectively.

A confidence measure of perhaps more practical use is an estimate of the probability that a given galaxy may have been classified as an elliptical when it is in reality a spiral. We thus calculate the fraction of objects within a given galaxy's bin that are classified as elliptical using the raw data but as spiral when the effect of the bias correction is taken into account. This fraction of misclassified galaxies, $f_{\rmn{misclass}}$, depends strongly on the threshold used in the analysis to define `spiral' or `elliptical', being highest when the \texttt{greater} condition is used (i.e. when a galaxy is classified as an elliptical when $\rmn{p_{el}>p_{sp}}$) and smaller when more stringent classifications are used.  However, the cost of using a more stringent classification threshold is that an increasing fraction of galaxies are unclassifiable, i.e. they do not meet the criteria for any of the classifications and are thus `uncertain'.  We quantify this by measuring the fraction of unclassified galaxies, $f_{\rmn{unclass}}$, in the same bin of redshift, magnitude and size as the particular galaxy in question.

\begin{figure*}
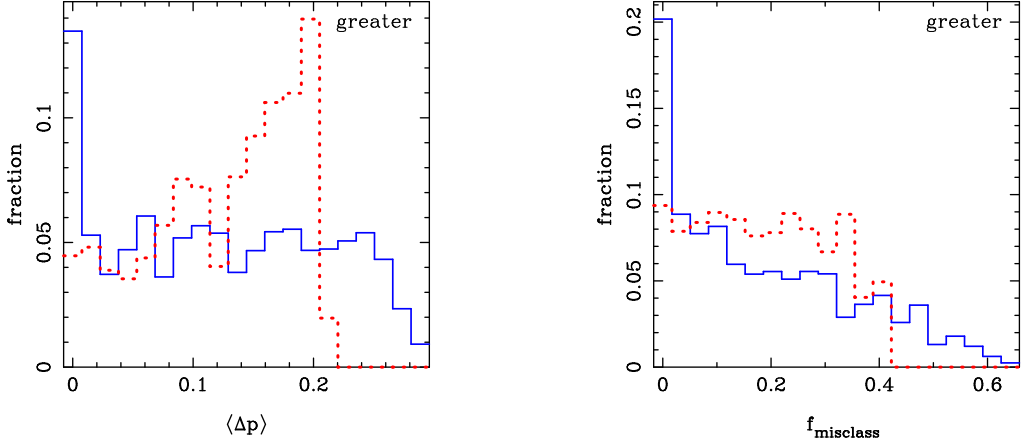

\centering
\hspace*{\stretch{1}}
\includegraphics[height=0.32\textwidth,angle=270]{check_confidence_greater_avcorr.ps}
\hspace*{\stretch{1}}
\includegraphics[height=0.32\textwidth,angle=270]{check_confidence_greater_misclass.ps}
\hspace*{\stretch{1}}
\caption{\label{fig:confgreater} Histograms showing
  \emph{(left)} the average bias correction applied to the
  type-likelihoods ($\Delta p$) and \emph{(right)} the estimated fraction of
  objects that are misclassified, at the absolute (blue, solid line)
  and apparent (red, dotted line) magnitude and size of each galaxy in
  the Galaxy Zoo Main Galaxy Sample.  These are calculated using the
  \texttt{greater} classification criteria, as described in the text.
  For example, note that $\sim 13.5$ per cent of galaxies are in absolute
  magnitude--size bins where approximately no bias correction is
  necessary, and $\sim 20$ per cent are in bins which have
  approximately no misclassified galaxies.  Conversely, $\sim 6$ per
  cent of galaxies are in bins where the bias correction changes the
  classifications for more than half of the objects.}
\end{figure*}

\begin{figure*}
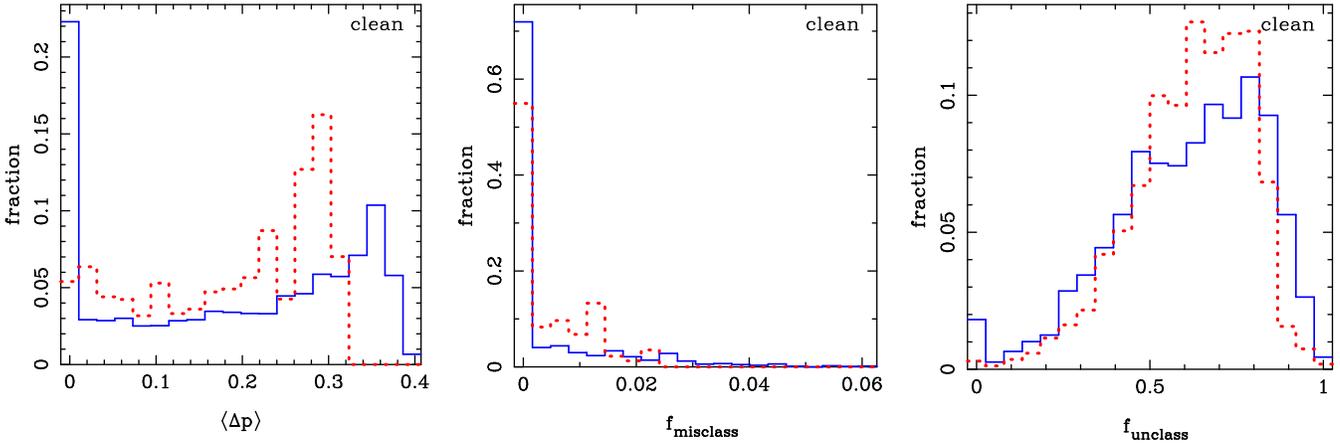

\centering
\includegraphics[height=0.32\textwidth,angle=270]{check_confidence_clean_avcorr.ps}
\hspace*{\stretch{1}}
\includegraphics[height=0.32\textwidth,angle=270]{check_confidence_clean_misclass.ps}
\hspace*{\stretch{1}}
\includegraphics[height=0.32\textwidth,angle=270]{check_confidence_clean_unclass.ps}
\caption{\label{fig:confclean} \emph{(Left)} and \emph{(centre)} as
  Fig. \ref{fig:confgreater}, but for galaxies classified using
the \texttt{clean} criteria.  This figure also includes \emph{(right)}
the fraction of objects that are unclassified by the \texttt{clean}
criteria, i.e. they have both $p_{\rmn{sp}}$ and $p_{\rmn{el}} < 0.8$.
The average corrections, are slightly larger for the \texttt{clean} versus
\texttt{greater} criteria, misclassifications are considerably lower,
but at the expense of a large fraction of unclassified galaxies. For example, over 70 percent of galaxies are in absolute magnitude-size bins for which the classification bias results in no objects being misclassified, but roughly two-thirds of galaxies are in absolute magnitude-size bins for which at least half the objects are unclassified.}
\end{figure*}

The distribution of these quantities amongst the Galaxy Zoo sample are shown in Figure \ref{fig:confgreater} (for the \texttt{greater} criterion) and Figure \ref{fig:confclean} (for \texttt{clean}). Note that weighting the results to favour those users who tend to agree with the majority makes little difference compared with the effect of changing the classification threshold.

\begin{figure*}
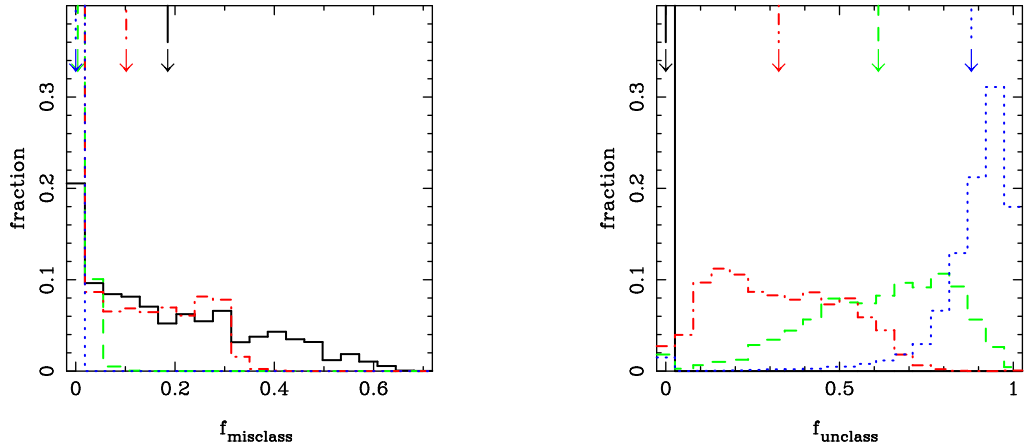

\centering
\hspace*{\stretch{1}}
\includegraphics[height=0.32\textwidth,angle=270]{check_confidence_misclass.ps}
\hspace*{\stretch{1}}
\includegraphics[height=0.32\textwidth,angle=270]{check_confidence_unclass.ps}
\hspace*{\stretch{1}}
\caption{\label{fig:conf_comparison} 
Histograms of \emph{(left)} fraction misclassified and \emph{(right)}
fraction unclassified, for \texttt{greater} (black, solid line),
\texttt{cleanish} (red, dot-dashed line), \texttt{clean} (green, dashed
line) and \texttt{superclean} (blue, dotted line).  The arrows at the
top indicate the means of each distribution.  One can clearly see that
as the classification criteria become more stringent the fraction of
misclassified objects decreases, but at the expenses of an increasing
fraction of unclassified objects.
}
\end{figure*}

Following earlier Galaxy Zoo papers, a \texttt{clean} sample has been defined by requiring 80 percent of the corrected vote to be in a particular category. However, this choice was somewhat arbitrary, and yet has a significant effect on the number of unclassified and misclassified galaxies. This effect is shown in Figure \ref{fig:conf_comparison} for four thresholds: 50 percent (\texttt{greater}), 60 percent (\texttt{cleanish}), 80 percent (\texttt{clean}) and 95 percent (\texttt{superclean}). The mean values of each distribution are indicated with arrows. For example, a threshold of 50 percent results, by design, in a classification for every galaxy but 19 percent are misclassified. A threshold of 60 percent results in 33 percent of galaxies unclassified and 10 percent misclassified. A threshold of 80 percent results in 60 percent of galaxies unclassified and 3 percent misclassified, while a threshold of 95 percent results in no misclassifications but 88 percent of galaxies unclassified. These figures illustrate the general principle of working with these data; as the the threshold is made more stringent, then the fraction of unclassified objects increases while the fraction of misclassified objects decreases. 

\section{The Catalogue}
\label{sec:classifications}

Table \ref{tab:maindata} contains the data for all MGS galaxies with measured redshifts in the range $0.001<z<0.25$ and $u$ and $r$ photometry in SDSS DR7, excluding those with extreme absolute magnitudes or sizes given by the SDSS pipeline. 667,945 galaxies are included. This table includes the raw votes, the weighted votes in elliptical (E) and combined spiral (CS) categories, and flags indicating the inclusion of the galaxy in a \texttt{clean}, debiased catalogue. The flags take into account not only the redshift dependence of the spiral/elliptical ratio as described in Section \ref{sec:bias} but also the redshift dependence of the ratio of spirals to ellipticals in the clean catalogue. This results in larger corrections than would otherwise be necessary. As explained above, bias correction was only possible for MGS galaxies for which SDSS DR7 included spectra, and so Table \ref{tab:nonspec} contains classifications for galaxies included in the Galaxy Zoo sample for which bias corrected morphologies are not available. 

As discussed in Section \ref{sec:prop}, while the introduction of mirrored and monochrome images was important in allowing the measurement of human bias, doing so also affected the behaviour of the participants. The measurements obtained during this bias study have thus not been combined with the main data set described above. Table \ref{tab:confabs}, presents the confidence measures discussed in section \ref{sec:confidence}, calculated using absolute magnitude and physical (rather than apparent) size. Tables \ref{tab:bias} and \ref{tab:bias2} gives details of the votes assigned to each category for the galaxies which were included in the bias study, as well as a combined vote in Table \ref{tab:combined}.

\subsection{Examples}

This paper presents, as a legacy for the community, the entire Galaxy Zoo 1 data set.  Users should bear in mind that the objects included in Galaxy Zoo 1 were selected by a combination of criteria (see \ref{sec:sample}), and therefore appropriate additional cuts (in magnitude, redshift, etc.) should be made to produce a well-defined sample appropriate to any particular study.

Most users of this data will have specific requirements which fall into one of a few categories. For example, one may require a small number of spiral or elliptical galaxies, perhaps for an observing proposal. In this situation, we suggest using a subset of the galaxies which we have flagged as belonging to the relevant category according to the \texttt{clean} criterion incorporating the bias-correction (Table \ref{tab:maindata}).

The data in Table \ref{tab:confabs} will then allow the user to estimate the fraction of the derived sample which are misclassified due to inherent classification bias (i.e. the inability to detect spiral arms in faint or small systems). The certainty of the individual classifications can be improved by using a higher threshold (e.g. requiring 95 per cent agreement amongst classifiers) or by selecting nearby, bright and/or large galaxies.

This procedure will suffice for many studies, but one should always consider the properties of the objects with `uncertain' classifications where these may potentially affect the result.  This will be the case for many statistical studies.  In such circumstances, it may be preferable to have an estimate of the morphology of all systems, rather than leaving a large number unclassified. In this case, the morphological type likelihoods from Table \ref{tab:maindata} (ideally the debiased quantities) may be used directly, or a simple majority vote can be applied.

Finally, if the small effect of the change in behaviour associated with the bias study can be ignored, and no bias-correction is required, a greater number of votes for $\sim 250,000$ systems can be obtained from Table \ref{tab:combined}. 

\section{Conclusions}

This paper presents the results of Galaxy Zoo 1, which used the World Wide Web to recruit a large community of volunteers to provide morphological classifications of galaxies drawn from the Sloan Digital Sky Survey. Such morphological classifications are useful indicators of a galaxy's dynamical state and are correlated with many other physical parameters. 

The data presented here has already produced several interesting results. Much of this work, published elsewhere (eg \citealp{Bamford, Schawinksiblue, Masters, Land}) was only possible because of the large number of morphological classifications provided by the project. The clockwise/anti-clockwise classifications of the spiral galaxies have been used to show that (as expected) there is no
evidence for a preferred rotation direction in the Universe, but the results suggest that
humans preferentially classify spiral galaxies as anti-clockwise
\citep{Land}. They hint at a local correlation of galaxy spins at
distances less than $\sim 0.5$Mpc - the first experimental evidence
for chiral correlation of spins \citep{Slosar}. 

The disentangling of
morphological and colour based classifications allows us to study
the separate dependences of morphology and colour on environment and
provide evidence that the transformation of galaxies from blue to red
proceeds faster than the transformation from spiral to early type (for example, \citealp{Bamford} and \citealp{Skibba}). The importance of this division is illustrated by the sample of passive, red, spirals in \citealp{Masters2}); these are disk galaxies in the outskirts of groups and clusters of galaxies which have either exhausted their gas, or lost it through strangulation or another mechanism. 

The Galaxy Zoo results can also be used to constrain the properties of dust in spiral galaxies \citep{Masters}. \citet{SchawinskiAGN} use Galaxy Zoo classifications to distinguish the host galaxies of AGN, finding that in the present day Universe activity is preferentially found in low mass early-types and high mass late-types. The sample of merging galaxies have been used to
show that the local fraction of mergers is between 1 and 3 percent and to study
the global properties of merging galaxies \citep{Darga, Dargb}.

Other serendipitous discoveries have been made because of the close attention given by classifiers to each image. Galaxy Zoo has
brought to light several rare classes of object. `Hanny's Voorwerp' - an
unusual emission line nebula neighbouring the spiral galaxy IC 2497
has been studied in several follow-up projects \citep{voorwerp,jozsa}. An
unusual class of emission line galaxies (the `peas') have been
discovered - their properties are discussed by \citet{peas} and \citet{amorin}.

The success of the project, both in quickly attracting large numbers of volunteers and in providing data that is useful for science, suggests that this mode of `citizen science' may provide a valuable method of data analysis for large data sets. A follow-up project, Galaxy Zoo 2\footnote{\url{http://zoo2.galaxyzoo.org}}, has obtained more than 60 million more detailed classifications of a subset of the Galaxy Zoo sample, and has already produced results; \citet{MastersZoo2} find that the presence of a bar is strongly linked to galaxy colour, with a bulge and bar-dominated sequence of red galaxies separated from a predominately bar-less blue cloud. 

Classification of galaxies drawn from large \emph{Hubble Space Telescope} surveys is now underway\footnote{\url{http://hubble.galaxyzoo.org}}. Two sister projects investigating transient detection\footnote{\url{http://supernova.galaxyzoo.org}} and merger simulations\footnote{\url{http://mergers.galaxyzoo.org}} are underway, and results from these projects will be reported in future papers. Data from a third spin-off, which asked users to determine the length of the bars in barred galaxies, is now being reduced\footnote{\url{
http://www.icg.port.ac.uk/~hoyleb/bars/}}. Obtaining a large number of visual classifications is not only inherently useful, but also provides a rich training set for improving automated techniques \citep{Banerji}; this combination of citizen science and machine learning will be essential in dealing with the data rates expected from future sky surveys, such as the Large Synoptic Survey Telescope \citep{LSST}. Whether used directly or to inform future surveys, Galaxy Zoo has shown that the efforts of volunteers, coordinated via the internet, can produce rich seams of science.

\section*{Acknowledgments}

The data in this paper is the result of the efforts of the Galaxy Zoo volunteers, without whom none of this work would have been possible. 

 Galaxy Zoo has been supported in part by a Jim Gray research grant from Microsoft, and by a grant from The Leverhulme Trust. CJL acknowledges support from the STFC Science in Society Program and The Leverhulme Trust, and thanks Prof. Joe Silk for his support. The team thank Jean Tate for assiduous copy-editing. This work is supported in part by the U.S. Department of Energy under Contract No. DE-AC02-98CH10886. Support for the work of K.S. was provided by NASA through Einstein Postdoctoral Fellowship grant number PF9-00069 issued by the Chandra X-ray Observatory Center, which is operated by the Smithsonian Astrophysical Observatory for and on behalf of NASA under contract NAS8-03060. KLM acknowledges funding from the Peter and Patricia Gruber
Foundation as the 2008 IAU Fellow, and from the University of
Portsmouth and SEPnet (www.sepnet.ac.uk). RCN thanks Google for partial funding during the Galaxy Zoo project. 

Funding for the SDSS and SDSS-II has been provided by the Alfred P. Sloan Foundation, the Participating Institutions, the National Science Foundation, the U.S. Department of Energy, the National Aeronautics and Space Administration, the Japanese Monbukagakusho, the Max Planck Society, and the Higher Education Funding Council for England. The SDSS Web Site is http://www.sdss.org/.

The SDSS is managed by the Astrophysical Research Consortium for the Participating Institutions. The Participating Institutions are the American Museum of Natural History, Astrophysical Institute Potsdam, University of Basel, University of Cambridge, Case Western Reserve University, University of Chicago, Drexel University, Fermilab, the Institute for Advanced Study, the Japan Participation Group, Johns Hopkins University, the Joint Institute for Nuclear Astrophysics, the Kavli Institute for Particle Astrophysics and Cosmology, the Korean Scientist Group, the Chinese Academy of Sciences (LAMOST), Los Alamos National Laboratory, the Max-Planck-Institute for Astronomy (MPIA), the Max-Planck-Institute for Astrophysics (MPA), New Mexico State University, Ohio State University, University of Pittsburgh, University of Portsmouth, Princeton University, the United States Naval Observatory, and the University of Washington.

\newpage
\clearpage

\tabletypesize{\scriptsize}
\begin{deluxetable}{lccccccccccccccc}
\tablecolumns{16}
\tablewidth{0pc}
\tablecaption{Classifications of MGS galaxies with spectra \label{tab:maindata}}
\tabletypesize{\scriptsize}
\rotate
\tablehead{
 & & & & \multicolumn{7}{|c|}{Votes$\rmn{^c}$}& \multicolumn{2}{|c|}{Debiased votes$\rmn{^d}$} &\multicolumn{3}{|c|}{Flags$\rmn{^e}$}\\
\colhead{ObjID$\rmn{^a}$} & \colhead{RA} & \colhead{Dec} & \colhead{$N_{\rmn{vote}}\rmn{^{b}}$} & \colhead{E} & \colhead{CW} & \colhead{ACW} & \colhead{Edge} & \colhead{DK} & \colhead{MG}  & \colhead{CS} & \colhead{E} & \colhead{CS} & \colhead{Spiral} & \colhead{Elliptical} & \colhead{Uncertain}\\
}
\small
\startdata
587727178986356823 & 00:00:00.41 & -10:22:25.7 & 59 & 0.61 & 0.034 & 0.0 & 0.153 & 0.153 & 0.051 & 0.186 & 0.61 & 0.186 & 0 & 0 & 1\\
587727227300741210 & 00:00:00.74 & -09:13:20.2 & 18 & 0.611 & 0.0 & 0.167 & 0.222 & 0.0 & 0.0 & 0.389 & 0.203 & 0.797 & 1 & 0 & 0\\
587727225153257596 & 00:00:01.03 & -10:56:48.0 & 68 & 0.735 & 0.029 & 0.0 & 0.147 & 0.074 & 0.015 & 0.176 & 0.432 & 0.428 & 0 & 0 & 1\\
587730774962536596 & 00:00:01.38 & +15:30:35.3 & 52 & 0.885 & 0.019 & 0.0 & 0.058 & 0.019 & 0.019 & 0.077 & 0.885 & 0.077 & 0 & 1 & 0\\
587731186203885750 & 00:00:01.55 & -00:05:33.3 & 59 & 0.712 & 0.0 & 0.0 & 0.22 & 0.068 & 0.0 & 0.22 & 0.64 & 0.29 & 0 & 0 & 1\\
587727180060098638 & 00:00:01.57 & -09:29:40.3 & 28 & 0.857 & 0.0 & 0.036 & 0.0 & 0.107 & 0.0 & 0.036 & 0.83 & 0.06 & 0 & 0 & 1\\
587731187277627676 & 00:00:01.86 & +00:43:09.3 & 38 & 0.5 & 0.0 & 0.053 & 0.289 & 0.105 & 0.053 & 0.342 & 0.351 & 0.473 & 0 & 0 & 1\\
587727223024189605 & 00:00:02.00 & +15:41:49.8 & 26 & 0.423 & 0.0 & 0.0 & 0.577 & 0.0 & 0.0 & 0.577 & 0.143 & 0.857 & 1 & 0 & 0\\
587730775499407375 & 00:00:02.10 & +15:52:54.2 & 62 & 0.355 & 0.016 & 0.21 & 0.323 & 0.0 & 0.097 & 0.548 & 0.355 & 0.548 & 0 & 0 & 1\\
587727221950382424 & 00:00:02.41 & +14:49:19.0 & 31 & 0.484 & 0.129 & 0.065 & 0.258 & 0.065 & 0.0 & 0.452 & 0.109 & 0.789 & 1 & 0 & 0\\
587730774425665704 & 00:00:02.58 & +15:02:28.3 & 24 & 0.583 & 0.042 & 0.125 & 0.167 & 0.083 & 0.0 & 0.333 & 0.147 & 0.701 & 0 & 0 & 1\\
587730773888794751 & 00:00:02.82 & +14:42:55.9 & 26 & 0.654 & 0.077 & 0.0 & 0.077 & 0.192 & 0.0 & 0.154 & 0.621 & 0.185 & 0 & 0 & 1\\
588015507658768464 & 00:00:03.24 & -01:06:46.8 & 57 & 0.474 & 0.088 & 0.0 & 0.263 & 0.175 & 0.0 & 0.351 & 0.324 & 0.48 & 0 & 0 & 1\\
587727178449485858 & 00:00:03.33 & -10:43:16.0 & 24 & 0.125 & 0.0 & 0.0 & 0.875 & 0.0 & 0.0 & 0.875 & 0.024 & 0.976 & 1 & 0 & 0\\
587730773351858407 & 00:00:03.46 & +14:11:53.6 & 64 & 0.625 & 0.016 & 0.016 & 0.25 & 0.078 & 0.016 & 0.281 & 0.245 & 0.597 & 0 & 0 & 1\\
587731187277693069 & 00:00:04.12 & +00:45:07.9 & 30 & 0.933 & 0.0 & 0.033 & 0.0 & 0.033 & 0.0 & 0.033 & 0.913 & 0.054 & 0 & 1 & 0\\
587727227837612116 & 00:00:04.18 & -08:44:03.0 & 37 & 0.73 & 0.0 & 0.081 & 0.135 & 0.054 & 0.0 & 0.216 & 0.648 & 0.295 & 0 & 0 & 1\\
587727225153257606 & 00:00:04.60 & -10:58:34.7 & 30 & 0.6 & 0.0 & 0.0 & 0.133 & 0.233 & 0.033 & 0.133 & 0.368 & 0.264 & 0 & 0 & 1\\
587727180596969574 & 00:00:04.60 & -08:56:37.6 & 30 & 0.167 & 0.333 & 0.033 & 0.467 & 0.0 & 0.0 & 0.833 & 0.075 & 0.925 & 1 & 0 & 0\\
587731187277693072 & 00:00:04.74 & +00:46:54.2 & 36 & 0.722 & 0.083 & 0.111 & 0.083 & 0.0 & 0.0 & 0.278 & 0.606 & 0.394 & 0 & 0 & 1\\
587727227300741221 & 00:00:05.17 & -09:13:04.6 & 66 & 0.424 & 0.0 & 0.03 & 0.061 & 0.03 & 0.455 & 0.091 & 0.335 & 0.133 & 0 & 0 & 1\\
588015507658768548 & 00:00:05.54 & -01:12:58.9 & 28 & 0.179 & 0.0 & 0.429 & 0.321 & 0.071 & 0.0 & 0.75 & 0.061 & 0.861 & 0 & 0 & 1\\
587727221413511423 & 00:00:05.70 & +14:24:44.8 & 56 & 0.321 & 0.036 & 0.339 & 0.143 & 0.125 & 0.036 & 0.518 & 0.069 & 0.721 & 0 & 0 & 1\\
587730775499407519 & 00:00:06.11 & +15:52:31.4 & 58 & 0.828 & 0.017 & 0.0 & 0.086 & 0.069 & 0.0 & 0.103 & 0.556 & 0.326 & 0 & 0 & 1\\
588015509806252152 & 00:00:06.67 & +00:30:16.8 & 38 & 0.711 & 0.0 & 0.0 & 0.132 & 0.158 & 0.0 & 0.132 & 0.612 & 0.216 & 0 & 0 & 1\\
587727221413511425 & 00:00:06.70 & +14:19:58.5 & 55 & 0.818 & 0.036 & 0.0 & 0.036 & 0.109 & 0.0 & 0.073 & 0.818 & 0.073 & 0 & 1 & 0\\
588015510343123099 & 00:00:07.12 & +00:51:28.5 & 47 & 0.766 & 0.021 & 0.043 & 0.149 & 0.021 & 0.0 & 0.213 & 0.602 & 0.373 & 0 & 0 & 1\\
587727220876640496 & 00:00:07.35 & +13:54:36.6 & 24 & 0.833 & 0.0 & 0.0 & 0.125 & 0.042 & 0.0 & 0.125 & 0.645 & 0.301 & 0 & 0 & 1\\
587730775499407506 & 00:00:07.37 & +15:51:19.2 & 31 & 0.742 & 0.0 & 0.065 & 0.0 & 0.097 & 0.097 & 0.065 & 0.513 & 0.183 & 0 & 0 & 1\\
587730775499407527 & 00:00:07.59 & +15:54:07.4 & 50 & 0.8 & 0.02 & 0.0 & 0.08 & 0.1 & 0.0 & 0.1 & 0.41 & 0.362 & 0 & 0 & 1\\
587730775499407394 & 00:00:07.62 & +15:50:03.2 & 31 & 1.0 & 0.0 & 0.0 & 0.0 & 0.0 & 0.0 & 0.0 & 1.0 & 0.0 & 0 & 1 & 0\\
\enddata
\tablecomments{The full version of this table is available from http://data.galaxyzoo.org. A portion is shown here for guidance regarding its form and content.}
\tablenotetext{a}{SDSS ID. This table includes all galaxies for which spectra are available in SDSS Data Release 7.}
\tablenotetext{b}{Total number of votes for each object.}
\tablenotetext{c}{Fraction of votes for Elliptical (E), ClockWise spirals (CW), AntiClockWise spirals (ACW), Edge-on spirals (Edge), Don't Know (DK), Merger (MG) and Combined Spiral (CS=Edge+CW+ACW) categories}
\tablenotetext{d}{Fraction of votes for Elliptical (E) and Combined Spiral (CS=Edge+CW+ACW) categories following the debiasing procedure described in section \ref{sec:bias}}
\tablenotetext{e}{Galaxies flagged as `elliptical' or `spiral' require 80 percent of the vote in that category after the debiasing procedure has been applied; all other galaxies are flagged `uncertain'.}
\end{deluxetable}

\newpage
\clearpage

\tabletypesize{\scriptsize}
\begin{deluxetable}{lcccccccccc}
\tablecolumns{11}
\tablewidth{0pc}
\tablecaption{Classifications of additional galaxies\label{tab:nonspec}}
\tabletypesize{\scriptsize}
\rotate
\tablehead{
& & & & \multicolumn{7}{|c|}{Votes$\rmn{^c}$}\\
\colhead{ObjID$\rmn{^a}$} & \colhead{RA} & \colhead{Dec} & \colhead{$N\rmn{^b}$} & \colhead{E} & \colhead{CW} & \colhead{ACW} & \colhead{Edge} & \colhead{DK} & \colhead{MG}  & \colhead{CS}\\
}
\small
\startdata
587730774425665700 & 00:00:01.28 & +15:04:40.8 & 73 & 0.479 & 0.0 & 0.0 & 0.014 & 0.479 & 0.027 & 0.014\\
587727220876640877 & 00:00:01.86 & +14:01:28.2 & 29 & 0.655 & 0.0 & 0.0 & 0.0 & 0.345 & 0.0 & 0.0\\
587727180060098742 & 00:00:02.15 & -09:31:37.0 & 30 & 0.467 & 0.0 & 0.033 & 0.0 & 0.467 & 0.033 & 0.033\\
588015509806252142 & 00:00:02.28 & +00:37:39.2 & 29 & 0.655 & 0.034 & 0.034 & 0.103 & 0.172 & 0.0 & 0.172\\
587731187277627683 & 00:00:02.96 & +00:43:04.8 & 24 & 0.583 & 0.0 & 0.083 & 0.167 & 0.125 & 0.042 & 0.25\\
587731186203951111 & 00:00:04.44 & -00:05:00.1 & 30 & 0.967 & 0.0 & 0.0 & 0.033 & 0.0 & 0.0 & 0.033\\
587730775499407505 & 00:00:04.96 & +15:51:15.3 & 45 & 0.667 & 0.0 & 0.044 & 0.178 & 0.044 & 0.067 & 0.222\\
587730774962536621 & 00:00:05.96 & +15:25:47.6 & 25 & 0.8 & 0.0 & 0.08 & 0.0 & 0.0 & 0.12 & 0.08\\
587730775499407504 & 00:00:07.22 & +15:51:14.2 & 39 & 0.59 & 0.0 & 0.026 & 0.026 & 0.051 & 0.308 & 0.051\\
587730773888794648 & 00:00:07.73 & +14:39:55.9 & 28 & 0.679 & 0.0 & 0.036 & 0.071 & 0.107 & 0.107 & 0.107\\
587727225690128558 & 00:00:08.42 & -10:28:23.6 & 43 & 0.512 & 0.023 & 0.0 & 0.442 & 0.023 & 0.0 & 0.465\\
587727222487318704 & 00:00:08.83 & +15:18:38.3 & 31 & 0.419 & 0.161 & 0.065 & 0.323 & 0.032 & 0.0 & 0.548\\
587727220876705929 & 00:00:09.64 & +14:05:42.8 & 25 & 0.28 & 0.04 & 0.04 & 0.36 & 0.28 & 0.0 & 0.44\\
587727225690128563 & 00:00:11.29 & -10:27:41.5 & 32 & 0.562 & 0.0 & 0.156 & 0.25 & 0.031 & 0.0 & 0.406\\
587727220876705935 & 00:00:11.92 & +14:05:24.0 & 31 & 0.258 & 0.032 & 0.0 & 0.0 & 0.097 & 0.613 & 0.032\\
587731187814563978 & 00:00:11.97 & +01:07:18.5 & 30 & 0.033 & 0.933 & 0.0 & 0.0 & 0.033 & 0.0 & 0.933\\
587730773351923943 & 00:00:13.06 & +14:13:18.0 & 31 & 0.903 & 0.032 & 0.0 & 0.065 & 0.0 & 0.0 & 0.097\\
587727177912615045 & 00:00:13.11 & -11:12:01.0 & 31 & 0.806 & 0.0 & 0.0 & 0.097 & 0.097 & 0.0 & 0.097\\
587730775499407560 & 00:00:14.32 & +15:52:16.7 & 29 & 0.448 & 0.379 & 0.0 & 0.103 & 0.069 & 0.0 & 0.483\\
587727179523227783 & 00:00:15.54 & -09:47:55.5 & 33 & 0.424 & 0.061 & 0.0 & 0.273 & 0.182 & 0.061 & 0.333\\
588015508732510387 & 00:00:16.12 & -00:13:58.4 & 22 & 0.545 & 0.0 & 0.182 & 0.091 & 0.182 & 0.0 & 0.273\\
588015508195639455 & 00:00:16.19 & -00:38:54.9 & 36 & 0.556 & 0.028 & 0.0 & 0.167 & 0.25 & 0.0 & 0.194\\
587727225153323052 & 00:00:17.96 & -10:53:39.8 & 54 & 0.556 & 0.056 & 0.056 & 0.167 & 0.148 & 0.019 & 0.278\\
588015508195639399 & 00:00:18.69 & -00:39:06.6 & 60 & 0.3 & 0.0 & 0.05 & 0.467 & 0.133 & 0.05 & 0.517\\
587727179523227873 & 00:00:20.63 & -09:48:34.5 & 52 & 0.538 & 0.0 & 0.0 & 0.058 & 0.115 & 0.288 & 0.058\\
587730774425665840 & 00:00:21.04 & +15:06:08.0 & 61 & 0.705 & 0.0 & 0.016 & 0.049 & 0.23 & 0.0 & 0.066\\
587730773351923979 & 00:00:21.19 & +14:10:53.7 & 33 & 0.212 & 0.0 & 0.0 & 0.152 & 0.061 & 0.576 & 0.152\\
587730774425665839 & 00:00:21.73 & +15:06:11.8 & 44 & 0.818 & 0.0 & 0.045 & 0.0 & 0.136 & 0.0 & 0.045\\
587727226763935920 & 00:00:22.65 & -09:38:18.6 & 52 & 0.327 & 0.115 & 0.115 & 0.365 & 0.077 & 0.0 & 0.596\\
588015507658768569 & 00:00:24.16 & -01:13:20.7 & 24 & 0.833 & 0.0 & 0.0 & 0.042 & 0.083 & 0.042 & 0.042\\
588015507658768511 & 00:00:24.17 & -01:14:44.9 & 32 & 0.875 & 0.031 & 0.031 & 0.031 & 0.031 & 0.0 & 0.094\\
587727226227064993 & 00:00:25.55 & -09:57:53.0 & 64 & 0.719 & 0.016 & 0.0 & 0.109 & 0.156 & 0.0 & 0.125\\
587727223024189787 & 00:00:25.61 & +15:41:28.2 & 33 & 0.879 & 0.0 & 0.03 & 0.03 & 0.061 & 0.0 & 0.061\\
\enddata
\tablecomments{The full version of this table is available from http://data.galaxyzoo.org. A portion is shown here for guidance regarding its form and content.}
\tablenotetext{a}{SDSS ID. This table includes all objects in the Galaxy Zoo sample for which spectra are not included in SDSS Data Release 7.}
\tablenotetext{b}{Total number of votes for each object.}
\tablenotetext{c}{Vote fractions for each category, as defined in the comments for Table \ref{tab:maindata}.}
\end{deluxetable}

\newpage
\clearpage

\tabletypesize{\scriptsize}
\begin{deluxetable}{lccccccccc}
\tablecolumns{12}
\tablewidth{0pc}
\tablecaption{Measures of confidence \label{tab:confabs}}
\tabletypesize{\scriptsize}
\rotate
\tablehead{
& & & \multicolumn{4}{|c|}{\texttt{clean}} & \multicolumn{3}{|c|}{\texttt{greater}}\\
\colhead{ObjID} & \colhead{RA} & \colhead{Dec} & \colhead{$f_{\rmn{unclass}}$} & \colhead{$f_{\rmn{misclass}}$} & \colhead{$\left< \Delta p \right>$} &\colhead{$\sigma_{\Delta p}$} & \colhead{$f_{\rmn{misclass}}$} & \colhead{$\left< \Delta p \right>$} &\colhead{$\sigma_{\Delta p}$}\\
}
\small
\startdata
587727178986356823 & 00:00:00.41 & -10:22:25.7 & 0.543 & 0.0 & 0.011 & 0.0090 & 0.0 & 0.0 & 0.0\\
587727227300741210 & 00:00:00.74 & -09:13:20.2 & 0.458 & 0.0 & 0.203 & 0.115 & 0.267 & 0.199 & 0.112\\
587727225153257596 & 00:00:01.03 & -10:56:48.0 & 0.811 & 0.046 & 0.367 & 0.192 & 0.2 & 0.174 & 0.09\\
587730774962536596 & 00:00:01.38 & +15:30:35.3 & 0.348 & 0.0 & 0.0 & 0.0 & 0.0 & 0.0 & 0.0\\
587731186203885750 & 00:00:01.55 & -00:05:33.3 & 0.83 & 0.0 & 0.302 & 0.122 & 0.053 & 0.055 & 0.026\\
587727180060098638 & 00:00:01.57 & -09:29:40.3 & 0.799 & 0.0 & 0.233 & 0.127 & 0.103 & 0.073 & 0.044\\
587731187277627676 & 00:00:01.86 & +00:43:09.3 & 0.852 & 0.0020 & 0.307 & 0.131 & 0.142 & 0.102 & 0.045\\
587727223024189605 & 00:00:02.00 & +15:41:49.8 & 0.75 & 0.0060 & 0.333 & 0.137 & 0.399 & 0.196 & 0.076\\
587730775499407375 & 00:00:02.10 & +15:52:54.2 & 0.629 & 0.0 & 0.035 & 0.019 & 0.0 & 0.0 & 0.0\\
587727221950382424 & 00:00:02.41 & +14:49:19.0 & 0.762 & 0.036 & 0.365 & 0.178 & 0.457 & 0.244 & 0.113\\
587730774425665704 & 00:00:02.58 & +15:02:28.3 & 0.774 & 0.027 & 0.375 & 0.177 & 0.529 & 0.261 & 0.117\\
587730773888794751 & 00:00:02.82 & +14:42:55.9 & 0.982 & 0.0 & 0.134 & 0.061 & 0.018 & 0.029 & 0.014\\
588015507658768464 & 00:00:03.24 & -01:06:46.8 & 0.869 & 0.0 & 0.0 & 0.0 & 0.216 & 0.107 & 0.045\\
587727178449485858 & 00:00:03.33 & -10:43:16.0 & 0.734 & 0.0060 & 0.333 & 0.126 & 0.517 & 0.228 & 0.08\\
587730773351858407 & 00:00:03.46 & +14:11:53.6 & 0.698 & 0.024 & 0.381 & 0.165 & 0.455 & 0.241 & 0.099\\
587731187277693069 & 00:00:04.12 & +00:45:07.9 & 0.527 & 0.0 & 0.0 & 0.0 & 0.011 & 0.045 & 0.028\\
587727227837612116 & 00:00:04.18 & -08:44:03.0 & 0.856 & 0.0 & 0.306 & 0.126 & 0.071 & 0.059 & 0.027\\
587727225153257606 & 00:00:04.60 & -10:58:34.7 & 0.797 & 0.046 & 0.374 & 0.191 & 0.303 & 0.197 & 0.097\\
587727180596969574 & 00:00:04.60 & -08:56:37.6 & 0.649 & 0.0 & 0.077 & 0.035 & 0.156 & 0.11 & 0.052\\
587731187277693072 & 00:00:04.74 & +00:46:54.2 & 0.336 & 0.0 & 0.046 & 0.042 & 0.034 & 0.042 & 0.039\\
587727227300741221 & 00:00:05.17 & -09:13:04.6 & 0.788 & 0.018 & 0.356 & 0.172 & 0.179 & 0.159 & 0.081\\
588015507658768548 & 00:00:05.54 & -01:12:58.9 & 0.674 & 0.0 & 0.034 & 0.017 & 0.183 & 0.127 & 0.074\\
587727221413511423 & 00:00:05.70 & +14:24:44.8 & 0.724 & 0.02 & 0.363 & 0.176 & 0.458 & 0.243 & 0.113\\
587730775499407519 & 00:00:06.11 & +15:52:31.4 & 0.797 & 0.046 & 0.374 & 0.191 & 0.303 & 0.197 & 0.097\\
588015509806252152 & 00:00:06.67 & +00:30:16.8 & 0.847 & 0.0 & 0.317 & 0.136 & 0.108 & 0.101 & 0.049\\
587727221413511425 & 00:00:06.70 & +14:19:58.5 & 0.604 & 0.0 & 0.0 & 0.0 & 0.0 & 0.0 & 0.0\\
588015510343123099 & 00:00:07.12 & +00:51:28.5 & 0.762 & 0.0080 & 0.351 & 0.147 & 0.132 & 0.11 & 0.05\\
587727220876640496 & 00:00:07.35 & +13:54:36.6 & 0.752 & 0.011 & 0.357 & 0.154 & 0.235 & 0.158 & 0.071\\
587730775499407506 & 00:00:07.37 & +15:51:19.2 & 0.503 & 0.0 & 0.204 & 0.12 & 0.322 & 0.214 & 0.128\\
587730775499407527 & 00:00:07.59 & +15:54:07.4 & 0.688 & 0.022 & 0.361 & 0.174 & 0.538 & 0.272 & 0.125\\
587730775499407394 & 00:00:07.62 & +15:50:03.2 & 0.167 & 0.0 & 0.0 & 0.0 & 0.0 & 0.043 & 0.046\\
\enddata
\tablecomments{These quantities are defined in section \ref{sec:confidence}.  The full version of this table is available from http://data.galaxyzoo.org. A portion is shown here for guidance regarding its form and content.}
\end{deluxetable}

\newpage
\clearpage

\tabletypesize{\scriptsize}
\begin{deluxetable}{lccccccccccccccccccc} 
\tablecolumns{25}
\tablewidth{0pc}
\tablecaption{Classification of mirrored images during bias study \label{tab:bias}}
\tabletypesize{\scriptsize}
\rotate
\tablehead{
& & & \multicolumn{8}{|c|}{Mirrored} & \multicolumn{8}{|c|}{Mirrored 2} \\
\colhead{ObjID$\rmn{^{a}}$} & \colhead{RA} & \colhead{Dec} & \colhead{$N_{\rmn{vote}}$} & \colhead{E} & \colhead{CW} & \colhead{ACW} & \colhead{Edge} & \colhead{DK} & \colhead{MG}  & \colhead{CS} & \colhead{$N_{\rmn{vote}}$} & \colhead{E} & \colhead{CW} & \colhead{ACW} & \colhead{Edge} & \colhead{DK} & \colhead{MG}  & \colhead{CS}\\
}
\footnotesize
\startdata
587727227300741210 & 00:00:00.74 & -09:13:20.2 & 111 & 0.396 & 0.135 & 0.0090 & 0.405 & 0.054 & 0.0 & 0.55 & 118 & 0.458 & 0.136 & 0.025 & 0.347 & 0.034 & 0.0 & 0.508\\
587727180060098638 & 00:00:01.57 & -09:29:40.3 & 121 & 0.876 & 0.0 & 0.017 & 0.033 & 0.066 & 0.0080 & 0.05 & 133 & 0.835 & 0.0 & 0.015 & 0.045 & 0.105 & 0.0 & 0.06\\
587727221950382424 & 00:00:02.41 & +14:49:19.0 & 132 & 0.629 & 0.015 & 0.083 & 0.197 & 0.068 & 0.0080 & 0.295 & 120 & 0.625 & 0.0 & 0.142 & 0.208 & 0.017 & 0.0080 & 0.35\\
587731186203951111 & 00:00:04.44 & -00:05:00.1 & 105 & 0.962 & 0.0 & 0.01 & 0.019 & 0.01 & 0.0 & 0.029 & 121 & 0.934 & 0.0 & 0.0 & 0.0 & 0.066 & 0.0 & 0.0\\
587731187814563978 & 00:00:11.97 & +01:07:18.5 & 122 & 0.025 & 0.033 & 0.91 & 0.0080 & 0.0080 & 0.016 & 0.951 & 126 & 0.0080 & 0.024 & 0.905 & 0.0080 & 0.032 & 0.024 & 0.937\\
587727225690128575 & 00:00:14.09 & -10:28:45.9 & 119 & 0.882 & 0.017 & 0.0080 & 0.067 & 0.025 & 0.0 & 0.092 & 100 & 0.94 & 0.01 & 0.0 & 0.02 & 0.03 & 0.0 & 0.03\\
587727223024189761 & 00:00:14.92 & +15:43:42.6 & 114 & 0.368 & 0.0 & 0.175 & 0.404 & 0.035 & 0.018 & 0.579 & 108 & 0.417 & 0.0090 & 0.102 & 0.398 & 0.074 & 0.0 & 0.509\\
588015508732510387 & 00:00:16.12 & -00:13:58.4 & 109 & 0.523 & 0.055 & 0.0 & 0.248 & 0.174 & 0.0 & 0.303 & 117 & 0.615 & 0.051 & 0.0 & 0.197 & 0.128 & 0.0090 & 0.248\\
587730773351923979 & 00:00:21.19 & +14:10:53.7 & 103 & 0.282 & 0.0 & 0.01 & 0.029 & 0.068 & 0.612 & 0.039 & 126 & 0.206 & 0.0080 & 0.0 & 0.079 & 0.071 & 0.635 & 0.087\\
588015507658768569 & 00:00:24.16 & -01:13:20.7 & 111 & 0.703 & 0.0090 & 0.018 & 0.054 & 0.09 & 0.126 & 0.081 & 119 & 0.731 & 0.0 & 0.0 & 0.067 & 0.101 & 0.101 & 0.067\\
587727223024189787 & 00:00:25.61 & +15:41:28.2 & 105 & 0.781 & 0.01 & 0.0 & 0.029 & 0.181 & 0.0 & 0.038 & 108 & 0.778 & 0.019 & 0.0 & 0.037 & 0.157 & 0.0090 & 0.056\\
587730773888794890 & 00:00:38.69 & +14:35:48.2 & 123 & 0.041 & 0.024 & 0.886 & 0.016 & 0.024 & 0.0080 & 0.927 & 116 & 0.043 & 0.0090 & 0.914 & 0.034 & 0.0 & 0.0 & 0.957\\
587727221950447853 & 00:00:38.70 & +14:53:40.8 & 107 & 0.421 & 0.0090 & 0.0 & 0.495 & 0.028 & 0.047 & 0.505 & 128 & 0.43 & 0.023 & 0.023 & 0.469 & 0.047 & 0.0080 & 0.516\\
587727221413577017 & 00:00:43.55 & +14:31:29.5 & 94 & 0.553 & 0.011 & 0.021 & 0.085 & 0.287 & 0.043 & 0.117 & 138 & 0.616 & 0.0070 & 0.014 & 0.094 & 0.261 & 0.0070 & 0.116\\
588015509269446865 & 00:00:46.32 & +00:03:54.9 & 140 & 0.943 & 0.0 & 0.0 & 0.036 & 0.014 & 0.0070 & 0.036 & 123 & 0.878 & 0.0080 & 0.0 & 0.041 & 0.065 & 0.0080 & 0.049\\
587727177912680582 & 00:00:47.50 & -11:06:12.7 & 127 & 0.756 & 0.0 & 0.0 & 0.087 & 0.11 & 0.047 & 0.087 & 113 & 0.823 & 0.027 & 0.0 & 0.0090 & 0.106 & 0.035 & 0.035\\
587727222487384226 & 00:00:50.19 & +15:21:44.2 & 128 & 0.836 & 0.0 & 0.0 & 0.016 & 0.031 & 0.117 & 0.016 & 108 & 0.843 & 0.0 & 0.0 & 0.028 & 0.046 & 0.083 & 0.028\\
587727222487384242 & 00:00:51.41 & +15:15:03.0 & 128 & 0.305 & 0.0 & 0.398 & 0.195 & 0.078 & 0.023 & 0.594 & 109 & 0.183 & 0.037 & 0.404 & 0.312 & 0.046 & 0.018 & 0.752\\
588015507658834049 & 00:00:51.75 & -01:11:53.9 & 128 & 0.914 & 0.0080 & 0.0 & 0.031 & 0.047 & 0.0 & 0.039 & 116 & 0.922 & 0.0 & 0.0090 & 0.052 & 0.017 & 0.0 & 0.06\\
587727225690259613 & 00:00:52.08 & -10:35:13.0 & 137 & 0.146 & 0.029 & 0.62 & 0.197 & 0.0070 & 0.0 & 0.847 & 120 & 0.092 & 0.025 & 0.75 & 0.133 & 0.0 & 0.0 & 0.908\\
588015507658833960 & 00:00:52.97 & -01:10:20.5 & 112 & 0.562 & 0.0 & 0.0 & 0.027 & 0.161 & 0.25 & 0.027 & 127 & 0.472 & 0.0 & 0.0080 & 0.031 & 0.181 & 0.307 & 0.039\\
588015507658834052 & 00:00:53.51 & -01:06:55.0 & 123 & 0.902 & 0.0080 & 0.033 & 0.033 & 0.024 & 0.0 & 0.073 & 118 & 0.958 & 0.0080 & 0.0 & 0.0080 & 0.025 & 0.0 & 0.017\\
587730775499473122 & 00:00:53.54 & +15:54:19.1 & 116 & 0.095 & 0.681 & 0.0 & 0.069 & 0.138 & 0.017 & 0.75 & 116 & 0.121 & 0.629 & 0.026 & 0.043 & 0.112 & 0.069 & 0.698\\
587727223561126046 & 00:00:56.33 & +16:05:33.7 & 118 & 0.831 & 0.0 & 0.017 & 0.059 & 0.085 & 0.0080 & 0.076 & 115 & 0.896 & 0.0 & 0.017 & 0.026 & 0.061 & 0.0 & 0.043\\
587727226764001349 & 00:01:05.80 & -09:42:40.8 & 111 & 0.892 & 0.0 & 0.0090 & 0.018 & 0.081 & 0.0 & 0.027 & 105 & 0.8 & 0.0 & 0.029 & 0.048 & 0.105 & 0.019 & 0.076\\
587727225690259639 & 00:01:06.37 & -10:24:00.9 & 134 & 0.037 & 0.791 & 0.03 & 0.112 & 0.0070 & 0.022 & 0.933 & 112 & 0.062 & 0.795 & 0.036 & 0.08 & 0.0090 & 0.018 & 0.911\\
587730774962602228 & 00:01:10.17 & +15:27:15.3 & 107 & 0.85 & 0.0 & 0.0 & 0.0090 & 0.14 & 0.0 & 0.0090 & 132 & 0.795 & 0.023 & 0.0080 & 0.061 & 0.114 & 0.0 & 0.091\\
587727223561191596 & 00:01:16.49 & +16:10:57.4 & 119 & 0.748 & 0.0 & 0.0 & 0.101 & 0.151 & 0.0 & 0.101 & 111 & 0.712 & 0.0 & 0.0090 & 0.063 & 0.216 & 0.0 & 0.072\\
587731185130340423 & 00:01:17.02 & -01:01:58.2 & 125 & 0.872 & 0.0080 & 0.0080 & 0.048 & 0.064 & 0.0 & 0.064 & 119 & 0.79 & 0.0080 & 0.0080 & 0.059 & 0.134 & 0.0 & 0.076\\
587727178986487929 & 00:01:26.64 & -10:11:51.9 & 115 & 0.8 & 0.0090 & 0.0 & 0.061 & 0.087 & 0.043 & 0.07 & 117 & 0.829 & 0.0090 & 0.0090 & 0.077 & 0.06 & 0.017 & 0.094\\
587727225690325075 & 00:01:31.48 & -10:28:54.5 & 107 & 0.888 & 0.0090 & 0.0 & 0.019 & 0.084 & 0.0 & 0.028 & 111 & 0.865 & 0.0 & 0.0 & 0.018 & 0.117 & 0.0 & 0.018\\
\enddata
\tablecomments{Classifications of galaxies during the bias study. Galaxies were shown mirrored about the vertical and diagonal axes (`Mirrored' and `Mirrored 2'). For each transformation we provide the total number of votes ($N_{\rmn{vote}}$) and vote fractions categories as defined in the comments for Table \ref{tab:maindata}. The full version of this table is available from http://data.galaxyzoo.org.}
\tablenotetext{a}{SDSS ID. This table includes all objects included in the Galaxy Zoo bias study sample, as described in section \ref{sec:prop}.}
\end{deluxetable}

\newpage
\clearpage
\tabletypesize{\scriptsize}
\begin{deluxetable}{lcccccccccc} 
\tablecolumns{25}
\tablewidth{0pc}
\tablecaption{Classification of monochrome images during bias study \label{tab:bias2}}
\tabletypesize{\scriptsize}
\rotate
\tablehead{
& & & \multicolumn{8}{|c|}{Monochrome}\\
\colhead{ObjID$\rmn{^{a}}$} & \colhead{RA} & \colhead{Dec} & \colhead{$N_{\rmn{vote}}$} & \colhead{E} & \colhead{CW} & \colhead{ACW} & \colhead{Edge} & \colhead{DK} & \colhead{MG}  & \colhead{CS}\\
}
\small
\startdata
587727227300741210 & 00:00:00.74 & -09:13:20.2 & 108 & 0.565 & 0.019 & 0.083 & 0.278 & 0.056 & 0.0 & 0.38\\
587727180060098638 & 00:00:01.57 & -09:29:40.3 & 123 & 0.854 & 0.0080 & 0.016 & 0.057 & 0.065 & 0.0 & 0.081\\
587727221950382424 & 00:00:02.41 & +14:49:19.0 & 137 & 0.467 & 0.117 & 0.029 & 0.321 & 0.066 & 0.0 & 0.467\\
587731186203951111 & 00:00:04.44 & -00:05:00.1 & 115 & 0.957 & 0.0 & 0.0 & 0.017 & 0.026 & 0.0 & 0.017\\
587731187814563978 & 00:00:11.97 & +01:07:18.5 & 120 & 0.033 & 0.883 & 0.025 & 0.0080 & 0.025 & 0.025 & 0.917\\
587727225690128575 & 00:00:14.09 & -10:28:45.9 & 108 & 0.907 & 0.0090 & 0.0 & 0.028 & 0.046 & 0.0090 & 0.037\\
587727223024189761 & 00:00:14.92 & +15:43:42.6 & 115 & 0.27 & 0.113 & 0.0 & 0.539 & 0.078 & 0.0 & 0.652\\
588015508732510387 & 00:00:16.12 & -00:13:58.4 & 111 & 0.649 & 0.0090 & 0.099 & 0.108 & 0.126 & 0.0090 & 0.216\\
587730773351923979 & 00:00:21.19 & +14:10:53.7 & 119 & 0.168 & 0.017 & 0.0 & 0.101 & 0.067 & 0.647 & 0.118\\
588015507658768569 & 00:00:24.16 & -01:13:20.7 & 117 & 0.598 & 0.06 & 0.0 & 0.094 & 0.197 & 0.051 & 0.154\\
587727223024189787 & 00:00:25.61 & +15:41:28.2 & 130 & 0.654 & 0.0 & 0.015 & 0.069 & 0.262 & 0.0 & 0.085\\
587730773888794890 & 00:00:38.69 & +14:35:48.2 & 97 & 0.041 & 0.918 & 0.01 & 0.0 & 0.021 & 0.01 & 0.928\\
587727221950447853 & 00:00:38.70 & +14:53:40.8 & 119 & 0.294 & 0.0080 & 0.017 & 0.613 & 0.042 & 0.025 & 0.639\\
587727221413577017 & 00:00:43.55 & +14:31:29.5 & 112 & 0.679 & 0.0 & 0.0090 & 0.062 & 0.214 & 0.036 & 0.071\\
588015509269446865 & 00:00:46.32 & +00:03:54.9 & 128 & 0.891 & 0.016 & 0.0 & 0.047 & 0.0080 & 0.039 & 0.062\\
587727177912680582 & 00:00:47.50 & -11:06:12.7 & 113 & 0.743 & 0.0 & 0.018 & 0.071 & 0.08 & 0.088 & 0.088\\
587727222487384226 & 00:00:50.19 & +15:21:44.2 & 104 & 0.808 & 0.0 & 0.01 & 0.01 & 0.048 & 0.125 & 0.019\\
587727222487384242 & 00:00:51.41 & +15:15:03.0 & 117 & 0.487 & 0.248 & 0.0090 & 0.171 & 0.077 & 0.0090 & 0.427\\
588015507658834049 & 00:00:51.75 & -01:11:53.9 & 121 & 0.868 & 0.0080 & 0.017 & 0.041 & 0.066 & 0.0 & 0.066\\
587727225690259613 & 00:00:52.08 & -10:35:13.0 & 120 & 0.242 & 0.592 & 0.017 & 0.117 & 0.033 & 0.0 & 0.725\\
588015507658833960 & 00:00:52.97 & -01:10:20.5 & 114 & 0.395 & 0.026 & 0.0 & 0.088 & 0.228 & 0.263 & 0.114\\
588015507658834052 & 00:00:53.51 & -01:06:55.0 & 98 & 0.98 & 0.01 & 0.0 & 0.01 & 0.0 & 0.0 & 0.02\\
587730775499473122 & 00:00:53.54 & +15:54:19.1 & 127 & 0.118 & 0.0 & 0.717 & 0.031 & 0.118 & 0.016 & 0.748\\
587727223561126046 & 00:00:56.33 & +16:05:33.7 & 96 & 0.771 & 0.021 & 0.01 & 0.042 & 0.156 & 0.0 & 0.073\\
587727226764001349 & 00:01:05.80 & -09:42:40.8 & 105 & 0.886 & 0.01 & 0.01 & 0.038 & 0.057 & 0.0 & 0.057\\
587727225690259639 & 00:01:06.37 & -10:24:00.9 & 118 & 0.059 & 0.017 & 0.831 & 0.068 & 0.0080 & 0.017 & 0.915\\
587730774962602228 & 00:01:10.17 & +15:27:15.3 & 110 & 0.882 & 0.0 & 0.0 & 0.027 & 0.091 & 0.0 & 0.027\\
587727223561191596 & 00:01:16.49 & +16:10:57.4 & 114 & 0.728 & 0.0090 & 0.0090 & 0.079 & 0.175 & 0.0 & 0.096\\
587731185130340423 & 00:01:17.02 & -01:01:58.2 & 114 & 0.798 & 0.0 & 0.018 & 0.07 & 0.114 & 0.0 & 0.088\\
587727178986487929 & 00:01:26.64 & -10:11:51.9 & 116 & 0.698 & 0.052 & 0.0090 & 0.078 & 0.129 & 0.034 & 0.138\\
587727225690325075 & 00:01:31.48 & -10:28:54.5 & 128 & 0.891 & 0.0080 & 0.016 & 0.031 & 0.055 & 0.0 & 0.055\\
\enddata
\tablecomments{Classifications of monochrome images of galaxies during the bias study. We provide the total number of votes ($N_{\rmn{vote}}$) and vote fractions in each of the categories, as defined in the comments for Table \ref{tab:maindata}. The full version of this table is available from http://data.galaxyzoo.org. A portion is shown here for guidance regarding its form and content.}
\tablenotetext{a}{SDSS ID. This table includes all objects included in the Galaxy Zoo bias study sample, as described in section \ref{sec:prop}.}
\end{deluxetable}

\newpage
\clearpage

\tabletypesize{\scriptsize}
\begin{deluxetable}{lcccccccccccccc}
\tablecolumns{15}
\tablewidth{0pc}
\tablecaption{Combined votes including bias study data \label{tab:combined}}
\tabletypesize{\scriptsize}
\rotate
\tablehead{
& & & & & & & & \multicolumn{7}{|c|}{Votes$\rmn{^{c}}$}\\
\colhead{ObjID$\rmn{^{a}}$} & \colhead{RA} & \colhead{Dec} &\colhead{$N_{\rmn{vote}}\rmn{^{b}}$} & \colhead{$N_{\rmn{vote,STD}}$} &  \colhead{$N_{\rmn{vote,MR1}}$} &  \colhead{$N_{\rmn{vote,MR2}}$} &  \colhead{$N_{\rmn{vote,MON}}$} &  \colhead{E} & \colhead{CW} & \colhead{ACW} & \colhead{Edge} & \colhead{DK} & \colhead{MG}  & \colhead{CS}\\
}
\small
\startdata
587727178986356823 & 00:00:00.41 & -10:22:25.7 & 59 & 59 & 0 & 0 & 0 & 0.61 & 0.034 & 0.0 & 0.153 & 0.153 & 0.051 & 0.186\\
587727227300741210 & 00:00:00.74 & -09:13:20.2 & 355 & 18 & 111 & 118 & 108 & 0.479 & 0.017 & 0.121 & 0.338 & 0.045 & 0.0 & 0.476\\
587727225153257596 & 00:00:01.03 & -10:56:48.0 & 68 & 68 & 0 & 0 & 0 & 0.735 & 0.029 & 0.0 & 0.147 & 0.074 & 0.015 & 0.176\\
587730774425665700 & 00:00:01.28 & +15:04:40.8 & 73 & 73 & 0 & 0 & 0 & 0.479 & 0.0 & 0.0 & 0.014 & 0.479 & 0.027 & 0.014\\
587730774962536596 & 00:00:01.38 & +15:30:35.3 & 52 & 52 & 0 & 0 & 0 & 0.885 & 0.019 & 0.0 & 0.058 & 0.019 & 0.019 & 0.077\\
587731186203885750 & 00:00:01.55 & -00:05:33.3 & 59 & 59 & 0 & 0 & 0 & 0.712 & 0.0 & 0.0 & 0.22 & 0.068 & 0.0 & 0.22\\
587727180060098638 & 00:00:01.57 & -09:29:40.3 & 405 & 28 & 121 & 133 & 123 & 0.854 & 0.012 & 0.0070 & 0.042 & 0.081 & 0.0020 & 0.062\\
587727220876640877 & 00:00:01.86 & +14:01:28.2 & 29 & 29 & 0 & 0 & 0 & 0.655 & 0.0 & 0.0 & 0.0 & 0.345 & 0.0 & 0.0\\
587731187277627676 & 00:00:01.86 & +00:43:09.3 & 38 & 38 & 0 & 0 & 0 & 0.5 & 0.0 & 0.053 & 0.289 & 0.105 & 0.053 & 0.342\\
587727223024189605 & 00:00:02.00 & +15:41:49.8 & 26 & 26 & 0 & 0 & 0 & 0.423 & 0.0 & 0.0 & 0.577 & 0.0 & 0.0 & 0.577\\
587730775499407375 & 00:00:02.10 & +15:52:54.2 & 62 & 62 & 0 & 0 & 0 & 0.355 & 0.016 & 0.21 & 0.323 & 0.0 & 0.097 & 0.548\\
587727180060098742 & 00:00:02.15 & -09:31:37.0 & 30 & 30 & 0 & 0 & 0 & 0.467 & 0.0 & 0.033 & 0.0 & 0.467 & 0.033 & 0.033\\
588015509806252142 & 00:00:02.28 & +00:37:39.2 & 29 & 29 & 0 & 0 & 0 & 0.655 & 0.034 & 0.034 & 0.103 & 0.172 & 0.0 & 0.172\\
587727221950382424 & 00:00:02.41 & +14:49:19.0 & 420 & 31 & 132 & 120 & 137 & 0.564 & 0.114 & 0.019 & 0.245 & 0.052 & 0.0050 & 0.379\\
587730774425665704 & 00:00:02.58 & +15:02:28.3 & 24 & 24 & 0 & 0 & 0 & 0.583 & 0.042 & 0.125 & 0.167 & 0.083 & 0.0 & 0.333\\
587730773888794751 & 00:00:02.82 & +14:42:55.9 & 26 & 26 & 0 & 0 & 0 & 0.654 & 0.077 & 0.0 & 0.077 & 0.192 & 0.0 & 0.154\\
587731187277627683 & 00:00:02.96 & +00:43:04.8 & 24 & 24 & 0 & 0 & 0 & 0.583 & 0.0 & 0.083 & 0.167 & 0.125 & 0.042 & 0.25\\
588015507658768464 & 00:00:03.24 & -01:06:46.8 & 57 & 57 & 0 & 0 & 0 & 0.474 & 0.088 & 0.0 & 0.263 & 0.175 & 0.0 & 0.351\\
587727178449485858 & 00:00:03.33 & -10:43:16.0 & 24 & 24 & 0 & 0 & 0 & 0.125 & 0.0 & 0.0 & 0.875 & 0.0 & 0.0 & 0.875\\
587730773351858407 & 00:00:03.46 & +14:11:53.6 & 64 & 64 & 0 & 0 & 0 & 0.625 & 0.016 & 0.016 & 0.25 & 0.078 & 0.016 & 0.281\\
587731187277693069 & 00:00:04.12 & +00:45:07.9 & 30 & 30 & 0 & 0 & 0 & 0.933 & 0.0 & 0.033 & 0.0 & 0.033 & 0.0 & 0.033\\
587727227837612116 & 00:00:04.18 & -08:44:03.0 & 37 & 37 & 0 & 0 & 0 & 0.73 & 0.0 & 0.081 & 0.135 & 0.054 & 0.0 & 0.216\\
587731186203951111 & 00:00:04.44 & -00:05:00.1 & 371 & 30 & 105 & 121 & 115 & 0.951 & 0.0030 & 0.0 & 0.013 & 0.032 & 0.0 & 0.016\\
587727180596969574 & 00:00:04.60 & -08:56:37.6 & 30 & 30 & 0 & 0 & 0 & 0.167 & 0.333 & 0.033 & 0.467 & 0.0 & 0.0 & 0.833\\
\enddata
\tablecomments{The full version of this table is available from http://data.galaxyzoo.org.}
\tablenotetext{a}{SDSS ID. This table includes all objects in the Galaxy Zoo bias sample.}
\tablenotetext{b}{Total number of votes for each object.}
\tablenotetext{c}{Vote fractions for each category, as defined in the comments for Table \ref{tab:maindata}.}
\end{deluxetable}

\label{lastpage}


\begin{thebibliography}{99}

\bibitem[\protect\citeauthoryear{Abraham, van den Bergh \& Nair}{2003}]{Abraham} Abraham, R.G., van den Bergh, S., Nair, P., 2003, ApJ, 588, 218


\bibitem[\protect\citeauthoryear{Adelman-McCarthy et al.}{2008}]{DR6} Adelman-McCarthy et al., 2008, ApJS, 175, 297 

\bibitem[\protect\citeauthoryear{Abazajian et al.}{2009}]{Abazajian} Abazajian, K.N. et al., 2009, ApJS, 182, 543

\bibitem[\protect\citeauthoryear{Amor\`in et al.}{2010}]{amorin} Amor\'in, P\'erez-Montero \& V\'ilchez, J.M., 2010, ApJL, 715,128


\bibitem[\protect\citeauthoryear{Ball et al.}{2004}]{Ball} Ball et al., 2004, MNRAS, 348, 1038

\bibitem[\protect\citeauthoryear{Banerji et al.}{2010}]{Banerji} Banerji et al., 2010, MNRAS, 406, 342

\bibitem[\protect\citeauthoryear{Bamford et al.}{2009}]{Bamford} Bamford, S.P. et al., 2009, MNRAS, 393, 1324 

\bibitem[\protect\citeauthoryear{Barbieri et al.}{2007}]{Barbieri} Barbieri et al., 2007, A\&A, 476, 13








\bibitem[\protect\citeauthoryear{Cardamone et al.}{2009}]{peas} Cardamone, C. et al., 2009, MNRAS, 399, 1191 

\bibitem[\protect\citeauthoryear{Conselice}{2006}]{Conselice} Conselice, C.J., 2006, MNRAS, 373, 1389

\bibitem[\protect\citeauthoryear{Darg et al.}{2010a}]{Darga} Darg, D., et al. 2010, MNRAS, 401, 1043

\bibitem[\protect\citeauthoryear{Darg et al.}{2010b}]{Dargb} Darg, D., et al., 2010, MNRAS, 401, 1552

\bibitem[\protect\citeauthoryear{Fukugita et al.,}{1996}]{Fukugita96} Fukugita, M., et al., 1996, AJ, 111, 1748

\bibitem[\protect\citeauthoryear{Fukugita et al.}{2007}]{Fukugita} Fukugita, M., et al., 2007, AJ, 134, 579

\bibitem[\protect\citeauthoryear{Hubble}{1936}]{Hubble} Hubble, E, 1936, \textit{The Realm of the Nebul\ae}, Yale University Press

\bibitem[\protect\citeauthoryear{J\'{o}zsa}{2009}]{jozsa} J\'{o}zsa et al., 2009, A\&A, 500, 33


\bibitem[\protect\citeauthoryear{Kinney et al.}{1996}]{Kinney} Kinney et al., ApJ, 467, 38

\bibitem[\protect\citeauthoryear{Lahav et al.}{1995}]{Lahav} Lahav, O. et al., 1995, Science, 267, 859

\bibitem[\protect\citeauthoryear{Land et al.}{2008}]{Land} Land, K. \& the Galaxy Zoo team 2008, MNRAS, 388, 1686


\bibitem[\protect\citeauthoryear{Lintott et al.}{2008}]{Lintott08} Lintott et al., 2008, MNRAS, 389, 1179

\bibitem[\protect\citeauthoryear{Lintott et al.}{2009}]{voorwerp} Lintott, C.J. et al., 2009, MNRAS, 399, 129


\bibitem[\protect\citeauthoryear{LSST Science Collaboration}{2009}]{LSST} LSST Science Collaboration, arXiv : 0912.0201


\bibitem[\protect\citeauthoryear{Masters et al.}{2010a}]{Masters} Masters, K., 2010a, MNRAS, 405, 783

\bibitem[\protect\citeauthoryear{Masters et al.}{2010b}]{Masters2} Masters, K., et al., 2010b, MNRAS, 404, 792

\bibitem[\protect\citeauthoryear{Masters et al.}{2010c}]{MastersZoo2} Masters K., et al., 2010c, Submitted to MNRAS., arXiv: 1003.0449

\bibitem[\protect\citeauthoryear{Mayo}{1933}]{Mayo} Mayo, E., 1933, The human problems of an industrial civilization (New York : MacMillian) ch.3

\bibitem[\protect\citeauthoryear{Mendez et al.}{2008}]{Mendez} Mendez, B. et al, ASP Conference Series vol. 389, ed. C. Garmany, M.Gibbs \& J.W. Moody, 219-226. 

\bibitem[\protect\citeauthoryear{Nair \& Abraham}{2010}]{Nair} Nair, P.B. \& Abraham, R.G., 2010, ApJS, 186, 427

\bibitem[\protect\citeauthoryear{Nieto-Santisteban, Szalay \& Gray}{2004}]{Nieto} Nieto-Santisteban, M., Szalay, A. \& Gray, J, 2004, in Astronomical Data Analysis \& Software Systems XIII (ASP Conference Series 314) eds F. Ochsenbein, M. Allen \& D. Egret, p 666



\bibitem[\protect\citeauthoryear{Roberts \& Haynes}{1994}]{Roberts} Roberts, M.S. \& Haynes, M.P., 1994, ARAA, 32, 115

\bibitem[\protect\citeauthoryear{Sandage}{1961}]{Sandage} Sandage, A.R., 1961, \textit{The Hubble Atlas of Galaxies}, Carnegie Institute of Washington, Washington

\bibitem[\protect\citeauthoryear{Sandage \& Visvanathan}{1978}]{Sandage2} Sandage, A.R. \& Visvanathan, N., 1978, ApJ, 225, 742


\bibitem[\protect\citeauthoryear{Schawinski et al.}{2007}]{Schawinski} Schawinski, K. et al., 2007, MNRAS, 382, 1415 

\bibitem[\protect\citeauthoryear{Schawinski et al.}{2009}]{Schawinksiblue} Schawinkski, K., \& the Galaxy Zoo team, 2009, MNRAS, 396, 818 

\bibitem[\protect\citeauthoryear{Schawinski et al.}{2010}]{SchawinskiAGN} Schawinski, K. et al., 2010, ApJ, 711, 284

\bibitem[\protect\citeauthoryear{Skibba et al.}{2009}]{Skibba} Skibba et al., 2009, MNRAS, 399, 966

\bibitem[\protect\citeauthoryear{Slosar et al.}{2009}]{Slosar} Slosar, A. et al., 2009, MNRAS, 392, 1225

\bibitem[\protect\citeauthoryear{Smith et al.}{2002}]{Smith} Smith, J.A. et al. 2002, AJ, 123, 2121

\bibitem[\protect\citeauthoryear{Strateva et al.}{2001}]{Strateva} Strateva et al., 2001, AJ, 122, 1861

\bibitem[\protect\citeauthoryear{Strauss et al.}{2002}]{Strauss} Strauss, M.A., et al., AJ, 123, 1810


\bibitem[\protect\citeauthoryear{Szalay et al.}{2002}]{Szalay02} Szalay, A. S., Gray, J., Thakar, A. R., Kunszt, P. Z., Malik, T., Raddick, M. J., Stoughton, C., \& vandenBerg, J. 2002, in Proceedings of the 2002 ACM SIGMOD International Conference on Management of Data, 570

\bibitem[\protect\citeauthoryear{de Vaucouleurs et al.}{1991}]{deVaucouleurs} de Vaucouleurs, G. et al., 1991, \textit{Third Reference Catalog of Bright Galaxies}, Springer-Verlag, New York. 

\bibitem[\protect\citeauthoryear{Westphal et al.}{2006}]{Westphal} Westphal A.J. et al., 2006, AGU, Fall Meeting, abstract P52B-08

\bibitem[\protect\citeauthoryear{York et al.}{2000}]{York} York, D.G. et al., 2000, AJ, 120, 1579

\end{thebibliography}
\end{document}